\documentclass[10pt, conference]{IEEEtran}
\usepackage{booktabs}
\usepackage{cite}
\usepackage{ulem}
\usepackage{graphicx}
\usepackage{caption3}
\usepackage{subfigure}
\usepackage[colorinlistoftodos]{todonotes}
\usepackage{indentfirst}
\usepackage{mathrsfs}
\usepackage{amssymb}
\usepackage{mathtools}
\usepackage{amsthm}
\usepackage{verbatim}
\usepackage{amsmath}
\usepackage{color}
\usepackage{listings}
\usepackage{diagbox}
\theoremstyle{plain}

\usepackage[linesnumbered,ruled]{algorithm2e}
\usepackage{array}
\usepackage{cite}
\def\BibTeX{{\rm B\kern-.05em{\sc i\kern-.025em b}\kern-.08em
		T\kern-.1667em\lower.7ex\hbox{E}\kern-.125emX}}
\makeatletter

\newcommand{\Rmnum}[1]{\expandafter\@slowromancap\romannumeral #1@}
\IEEEoverridecommandlockouts
\usepackage{cite}
\usepackage{amsmath,amssymb,amsfonts}
\usepackage{algorithmic}
\usepackage{graphicx}
\usepackage{textcomp}
\usepackage{xcolor}
\def\BibTeX{{\rm B\kern-.05em{\sc i\kern-.025em b}\kern-.08em
    T\kern-.1667em\lower.7ex\hbox{E}\kern-.125emX}}
\begin{document}

\title{Wireless Edge-Empowered Metaverse: A Learning-Based Incentive Mechanism for Virtual Reality}
\author{Minrui~Xu, Dusit~Niyato, Jiawen~Kang, Zehui~Xiong, Chunyan~Miao, and Dong~In~Kim
    \thanks{M.~Xu, D.~Niyato, and C.~Miao are with the School of Computer Science and Engineering, Nanyang Technological University, Singapore (e-mail: minrui001@e.ntu.edu.sg; dniyato@ntu.edu.sg; ascymiao@ntu.edu.sg)}
    \thanks{J.~Kang is with the School of Automation, Guangdong University of Technology, Guangzhou 510006, China (e-mail: kavinkang@gdut.edu.cn)}
  \thanks{Z.~Xiong is with the Pillar of Information Systems Technology and Design, Singapore University of Technology and Design, Singapore 487372, Singapore (e-mail: zehui\_xiong@sutd.edu.sg)}
  \thanks{D. I. Kim is with the Department of Electrical and Computer Engineering, Sungkyunkwan University, Suwon 16419, South Korea (e-mail: dikim@skku.ac.kr)}
 }
 

\maketitle

\begin{abstract}
The Metaverse is regarded as the next-generation Internet paradigm that allows humans to play, work, and socialize in an alternative virtual world with immersive experience, for instance, via head-mounted display for Virtual Reality (VR) rendering. With the help of ubiquitous wireless connections and powerful edge computing technologies, VR users in the wireless edge-empowered Metaverse can immerse themselves in the virtual through the access of VR services offered by different providers. However, VR applications are computation- and communication-intensive. The VR service providers (SPs) have to optimize the VR service delivery efficiently and economically given their limited communication and computation resources. An incentive mechanism can be thus applied as an effective tool for managing VR services between providers and users. Therefore, in this paper, we propose a learning-based Incentive Mechanism framework for VR services in the Metaverse. First, we propose the quality of perception as the metric for VR users immersing in the virtual world. Second, for quick trading of VR services between VR users (i.e., buyers) and VR SPs (i.e., sellers), we design a double Dutch auction mechanism to determine optimal pricing and allocation rules in this market. Third, for auction information exchange cost reduction, we design a deep reinforcement learning-based auctioneer to accelerate this auction process. Experimental results demonstrate that the proposed framework can achieve near-optimal social welfare while reducing at least half of the auction information exchange cost than baseline methods.
\end{abstract}


\section{Introduction}
	%
	%
	%
	%
	The Metaverse is the next-generation Internet after the web and the mobile network revolutions, in which humans (acting as digital avatars) can interact with other people and software applications in a  three-dimensional (3D) virtual world~\cite{10.1145/3474085.3479238, all2021Lee}. As described in the famous film \textit{Ready Player One}, the Metaverse allows everyone acting as a customized avatar to be connected and do everything at will only following some basic rules. The Metaverse provides users with immersive experiences in the virtual world by head-mounted display (HMD) for Augmented Reality (AR) and Virtual Reality (VR)~\cite{chen2018virtual}. As envisioned in \textit{Ready Player One}, the Metaverse is feasible under today's reachable technologies, such as digital twins for modeling of the physical world, VR/AR for immersive experience via HMD rendering, and Blockchain for distributed ledgers and decentralized autonomous organizations (DAOs)~\cite{10.1145/3474085.3479238}. Empowered by 5G and beyond wireless networks and edge computing/intelligence technologies, the Metaverse allows users to access this virtual world anywhere and anytime. In the wireless edge-empowered Metaverse, VR users can feel and experience seamlessly immersive in various Metaverse applications, such as massively multiplayer online (MMO) videogames and virtual concerts, with ultra-low latency and ultra-high reliable connections~\cite{chen2018virtual}.
    
   Although wireless edge-empowered Metaverse has so many advantages, its economic system design (especially the incentive mechanisms) for motivating and incentivizing both users and service providers (SPs) to participate in the Metaverse interaction still needs much further investigation. For example, when playing MMO video games, VR users need to allocate a large amount of communication and computation resources from SPs to guarantee their quality of experience. However, there are three main problems in designing economic mechanisms in the wireless edge-empowered Metaverse. First, unlike omnidirectional VR applications, wireless edge-empowered Metaverse renders the virtual world via non-panoramic VR for the communication traffic cost and the computation consumption reductions of wireless devices~\cite{10.1145/3394171.3413681}. In this context, the utility of VR users and SPs in the Metaverse needs to be redesigned to emphasize the quality of perception experience of VR users~\cite{niyato_luong_wang_han_2020}. Second, the Metaverse service market is a bilateral market with VR users acting as buyers and SPs acting as sellers, thus complicating their interactions. Third, unlike continuous trading markets, the Metaverse service market is a call market and requires efficient matching and pricing between VR users and SPs~\cite{10.2307/2234848}.
   
    To address these problems, the Double Dutch Auction (DDA)~\cite{friedman2018double}, a generalization of Dutch auction in the double-side market, is a promising solution. The auctioneer in DDA, run by DAOs in the Metaverse~\cite{10.1145/3474085.3479238}, uses two Dutch clocks to match and determine transactions for buyers (i.e. VR users) and sellers (i.e., VR SPs) in the VR service market. The Buyer clock, starting from high and descending over time, shows the current buy price of VR services, and the Seller clock, starting from low and ascending over time, shows the current sell price of VR services. The bidding process of DDA is asynchronous and iterative. In each auction round, the auctioneer first broadcasts the current auction clocks to buyers and sellers, respectively.  The buyers and sellers indicate their willingness to buy and sell and stop their clocks at a market price at least as favorable as their respective clock prices. During the auction, the clocks are always restarted at the same value at which they were stopped. This allows multiple units to be traded at the same clock price. Finally, the auction ends when the two clocks cross and the auctioneer clears the market at the common crossing price between the winning buyers and winning sellers. Furthermore, via Deep Reinforcement Learning (DRL)~\cite{schulman2017proximal}, the auctioneer can learn to adjust the clock stepsize of the auction clocks dynamically during the auction process without prior auction knowledge. In this way, this learning-based auctioneer can maximize social welfare while improving auction efficiency.
	
	In this paper, we propose an efficient deep reinforcement learning-based incentive mechanism framework to manage and allocate non-panoramic VR services between SPs and users in wireless edge-empowered Metaverse. For efficient determination of the allocation and pricing rules for VR services, we propose the DDA call market, in which the auction is completed in a finite time. In order to improve the efficiency of DDA, that is, to reduce the auction information exchange cost within the VR service market, the auctioneer acting as the agent in deep reinforcement learning can learn the near-optimal policy through interaction with the environment without prior auction knowledge. 
	
	Our contributions can be summarized as follows:
	\begin{itemize}
	    \item We present an efficient learning-based incentive mechanism framework for wireless edge-empowered Metaverse to evaluate and reinforce the seamless experience of immersion of VR users, which is the first market mechanism for VR services in the Metaverse. 
        \item Unlike existing double auctions for continuous trading markets, we propose the double Dutch auction to match and price  VR services between users and VR SPs dynamically for our framework. The SPs and users can offer their buy-bids and sell-bids asynchronously but trade simultaneously while guaranteeing incentive rationality, truthfulness, and budget balance.
        \item To improve the DDA efficiency, we design a DRL-based DDA to maximize social welfare with high communication efficiency. The auctioneer act as the DRL agent to learn the efficient auction policy without prior knowledge and reduce the auction information exchange cost.
	\end{itemize}
    \section{Wireless Edge-empowered Metaverse Model}

	We consider a wireless edge-empowered Metaverse system with a set $\mathcal{N} = \{1, \ldots, n, \ldots, N\}$ of $N$ VR SPs and a set $\mathcal{M} = \{1, \ldots, m, \ldots, M\}$ of $M$ VR users in a downlink wireless network \cite{niyato_luong_wang_han_2020}. As shown in Fig.1, the wireless edge-empowered Metaverse focuses on providing non-panoramic VR applications for interactive entertainments, e.g., immersive videogames and virtual concerts, through wireless connections~\cite{Kelkkanen2020BitrateRO}. By the non-panoramic VR, users can immerse in the Metaverse with 3D and high-resolution video steams with 3D surround sound from SPs. The SPs, which render 3D scenes for VR applications, first encode the resulting content using the NVENC H.264 codes. The SPs transmit them to their VR users via wireless networks. Finally, the VR users decode and display the content in their HMD devices.
	
	\subsection{Communication Model}
	
	In the Metaverse, maintaining high bitrate among VR users is a desiring property for immersive experience. However, the solution for streaming of 360 VR by using pre-rendered panorama images requires on average bitrate at 132 Mbps~\cite{Kelkkanen2020BitrateRO}. Fortunately, the bitrate recommendation to support non-panoramic VR for the HMD resolution is a range of 20-38 Mbps. This is also a heavy burden on wireless networks, so we consider an frequency division multiple access (FDMA) system that each SP has one resouce block to server one VR user, for a more general case. Accordingly, the downlink bitrate for the VR user under SP $n$ is described as\cite{chen2018virtual}:
    \begin{equation}
    R_n = B_n\mathbb{E}_{h_n}\left(\log_2(1+\frac{p_n|h_n|^2}{B_nN_0})\right),
    \end{equation}where $B_n$ denotes the allocated bandwidth of SP $n$ in its resource block, $p_n$ refers to the transmit power of SP $n$, $N_0$ is the noise power spectral density, and $\mathbb{E}_{h_n}(\cdot)$ refers to the expectation with respect to the channel gain $h_n$ between VR SP $n$ and the VR user. Similar to~\cite{9562560}, we consider that each orthogonal resource block allocated for the VR service is occupied by at most one VR user, namely, the interference caused by other VR SPs or users participating in the wireless edge-empowered Metaverse could be ignore.
		\begin{figure}[!]
		\vspace{-0.185cm}
		\centering
		\includegraphics[width=1\linewidth]{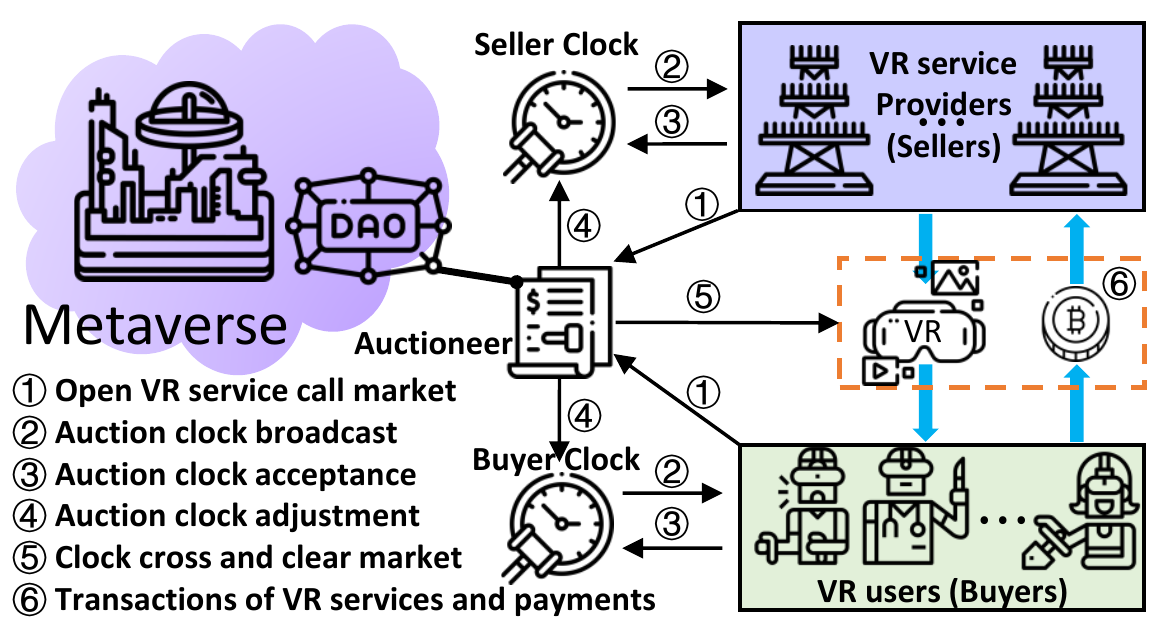}
		\caption{A Typical Framework of Wireless Edge-empowered Metaverse.}
		\label{fig_system}
		\vspace{-0.185cm}
	\end{figure}
    
	\subsection{Quality of Perception Experience Model for the Metaverse}
	
	Different from omnidirectional VR, which requires images to cover a much larger area, non-panoramic VR demands less data traffic and rendering power to support similar quality of perception experience of VR applications. However, non-panoramic VR requires different metrics to evaluate the quality of perception experience of VR users. Continuous Rotation takes ten samples of instantaneous Structural SIMilarity (SSIM)~\cite{wang2004image} and Video Multi-Method Assessment Fusion (VMAF)~\cite{Orduna2020VideoMA} within a constant rotation at various bitrates. SSIM measures the similarities between two images in non-panoramic VR rendering. Based on previous work, the authors in \cite{Cuervo2015DemoKH}, used the number, 0.98, as the limit for a visually lossless experience. VMAF is a perceptual video quality assessment algorithm developed by Netflix seeking to reflect the viewer’s perception of our streaming quality. Although VMAF is initially designed to work with traditional 2D content, it can also be successfully used without any specific adjustments to obtain the quality of VR sequences actually perceived by users. The authors in~\cite{Orduna2020VideoMA} provide extensive experimental tests on omnidirectional VR content.
	Accordingly, we formulate our quality of VR perception experience model based on SSIM and VMAF.
	Let $V_m$ be the average rotation speed of VR user $m$ and $R_m$ be the required bitrate to support its VR applications. Due to the constrictions of human health and fierce online games, the average rotation speed is between 360$^{\circ}$ and 1080$^{\circ}$, i.e., 360$^{\circ}$ $\leq V_m \leq$ 1080$^{\circ}$. The SSIM perceived by VR user $m$ is given by~\cite{Kelkkanen2020BitrateRO}
	\begin{equation}
		f(V_m,R_m) = \max\{b_0,1-(b_1+b_2V_m)R_m^{-(b_3+b_4V_m)}\},
	\end{equation}
	with coefficients $b_i$ listed in Table I. As $V_m$ grows, SSIM values approach the optimal value 1 in an asymptotic way. Also, with the corresponding fits for $V_m$, VMAF for VR user $m$ can be expressed as~\cite{Kelkkanen2020BitrateRO}
	\begin{equation}
		g(V_m,R_m)=\min\{100,b_1+b_2V_m+b_3R_m+b_4V_mR_m\},
	\end{equation}
	where $b_i$ are coefficients listed in Table I. The negative sign of coefficients $b_2$ and $b_4$ indicates that as $R_m$ increases, the value of $V_m$ slows down to approach the maximum of 100. VMAF values, however, increase earlier and steeper as $R_m$ increases from their minimal values, finally reaching their peak value of 100, unlike SSIM.
	
    \begin{table}[!]
    \caption{Coefficients of the SSIM and the VMAF formulates~\cite{Kelkkanen2020BitrateRO}}
    \small\centering
        \begin{tabular}{|c|c|c|c|c|}
        \hline
             & $b_1$  & $b_2$       & $b_3$      & $b_4$       \\ \hline
        SSIM & 0.65  & 0.368    & 1.23e-3 & 0.850    \\ \hline
        VMAF & 36.13 & -1.66e-2 & 11.62   & -6.07e-3 \\ \hline
        \end{tabular}
        \vspace{-4mm}
    \end{table}
    
	\section{Market Design and Problem Formulation}

	\subsection{VR Service Call Market}
	
	In the security industries, orders are batched for simultaneous execution in multiple-unit auctions in what is referred to as ``call markets"; that is, the security is ``called" for auction at a particular point in time. This type of market is used on the stock exchanges in many countries~\cite{friedman2018double}. 
	
	As the wireless VR applications, such as MMO video games and virtual concerts, should be loaded in a short period of time. The VR service market in the wireless edge-empowered Metaverse is a kind of call market, in which VR SPs and users open and finish a VR service trade instantaneously.
	At the loading process of the games or concerts, VR users and VR SPs open a VR service call market, as shown in Fig. 1, where VR users are the buyers and SPs are the sellers, to trade VR services under the supervision of DAO, acting as auctioneer. Let $R^{req}$ be the required bitrate for VR users to run the VR applications. However, the provision of the VR services satisfying $R^{req}$ needs to jointly consider the communication resources and the computation resources of SPs. For example, the value of communication resources depends on the spectrum bands which vary due to the time-varying demand~\cite{niyato_luong_wang_han_2020}. The valuation of computation resources depends on the CPU frequencies of SPs~\cite{chen2018virtual}. Thus, the SP $n$ determines its valuation function $v_n^s$ on its resource and the rendering requirements, in terms of resolution $D$ and frame rate $f$, of the VR services, which can be expressed as
	\begin{equation}
	    v_n^s = Df(w_1 E_n B_n+ w_2 \alpha_nf_n^2),
	\end{equation}
	where $E_n$ denotes the spectrum efficiency of SP $n$ determined by the communication model in (1), $\alpha_n$ denotes the effective capacitance coefficient of SP $n$’s computing chipset, the $f_n$ is its CPU frequency, and $w_1+w_2=1$ are weights.

	The valuation function $v_m^b$ of buyer $m$ for the VR service depending on its average rotation speed and expected bitrate can be calculated as
	\begin{equation}
		v_m^b(V_m, R_m) = g(V_{m},R_m)^{f(V_{m},R_m)},
	\end{equation}
	where $V_m$ is the average rotation speed of user $m$ and $R_m$ is the expected bitrate of user $m$.

	\subsection{Double Dutch Auction for VR Services}
	
	Considering multiple VR SPs and users in the wireless edge-empowered Metaverse, we utilize the DDA mechanism for the VR service double-side call market. The objectives of the double auction mechanism are (i) to match the SPs to VR users, and (ii) to determine the buying prices of VR users and selling prices of SPs.
	
	In the VR service call market, there are several entities acting in different roles as follows:
	
	\subsubsection{VR users (Buyers)}
	The VR users are the buyers in this call market that pay the SPs for their resources in facilitating the VR applications of users. In order to participate in the auction, the VR users submit their buy-bid to the auctioneer. The buy-bid of VR users $m$ at time slot $t$, which is denoted by $k_m^t$, represents the maximum price that VR user $m$ is willing to pay for the VR service from SPs, which can be calculated as $k_m^t = v_m^b(V_m,R^{req})$,
where $R^{req}$ is the requried bitrate for the VR service to support current VR applications. The VR users are able to decide on the average rotation speed and the acquired bitrate of their VR applications.
	\subsubsection{VR SPs (Sellers)}
    The VR SPs are the sellers that provide their computation and communication resources to deliver VR services for users. In return, they receive monetary rewards for the provision of their resources. By participating in the DDA mechanism, the sellers submit their offering sell-bids to the auctioneer, which can be calculated as $o_n^{t} = v_n^s$.
	\subsubsection{DAO (Auctioneer)}
	The auctioneer of DDA is run by the DAOs in the Metaverse. The auctioneer of DDA maintains two Dutch clocks for the buyer side $\Psi^t=0$ and the seller side $\Psi^t=1$ simultaneously, where $\Psi^t$ is the flag denoting the current trading side. The buyer clock $C_B$ shows the buying price which opens with the highest price $C_B^0 = p^{max}$ and descends over time. On the other hand, the seller clock $C_S$ shows the selling price which opens with the lowest price $C_S^0 = p^{min}$ and ascends over time. The auctioneer usually starts with the buyer round, i.e., $\Psi^0 = 0$. 

Without loss of generality,  we assume that the buyers are arranged according to their values non-increasingly, i.e., $v_1^b\geq \cdots \geq v_m^b\geq \cdots\geq v_M^b$. And the sellers are arranged according to their value non-decreasingly, i.e., $v_1^s\leq  \cdots\leq v_n^s \leq\cdots\leq v_N^s$. There are three steps in the DDA, i.e., Clock Broadcast, Clock Acceptance, and Clock Adjustment. The following terminologies are used in the DDA:

\begin{enumerate}
\item Clock Broadcast: At the beginning of each auction round, the auctioneer broadcasts the current clock to buyers or sellers in the market. When $\Psi =0$, which means the auction is on the buyer's side, the auctioneer broadcasts the current buyer clock $C_B^t$ to all the buyers. If the auction is on the seller side, i.e., $\Psi=1$, the current seller clock $C_S^t$ is sent to all the sellers.
\item Clock Acceptance: After receiving the auction clock from the auctioneer, the buyers and sellers compare their bids with their current bids. If the current auction clock meets the bid conditions, the buyers or sellers whose conditions are met will bid or ask the auctioneer. Specifically, when $\Psi^t=0$, which means the current auction is on the buyer side, the $m$-th available buyer, i.e., $m\notin\mathcal{M}_B^{t}$, submits its buy-bid as the current buyer clock $b_m^t \leftarrow C_B^{t}$ if it accepts the clock $k_m^t \geq C_B^{t}$. Then the set of winning buyers adds the buyer $m$ in current trade $\mathcal{M}_B^{t+1} = \mathcal{M}_B^{t}\cup \{m\}$. Thus, the difference between the expected buy-bid and the real buy-bid is the regret $e_m^t$ of buyer $m$, which can be calculated as
	\begin{equation}
		e_m^t = k_m^b - b_m^t.
	\end{equation}
	After that, the auctioneer will change the flag $\Psi^{t+1}$ to 1 and turn to the seller side. If the current auction market is on the seller side, i.e., $\Psi^t=1$, the situation is similar. The sellers who have not yet traded will judge whether they want to participate in this round of transactions after receiving the seller's clock from an auctioneer. If the conditions for participating in the transaction are met, that is, $o_n^t\leq C_S^t$ and $n\notin\mathcal{N}_S^{t}$, the seller will accepts its sell-bid to the current seller clock, i.e., $a_n^t \leftarrow C_S^t$ to inform auctioneer that it wants to trade. The auctioneer will then add the seller to set of winning sellers, i.e., $\mathcal{N}_S^{t+1}=\mathcal{N}_S^{t}\cup\{n\}$. Therefore, the difference between the expected sell-bid and the real sell-bid is the regret $e_n^t$ of seller $n$, which can be calculated as
	\begin{equation}
	e_n^t = a_n^t - o_n^{t}.
	\end{equation}
	The auctioneer then adjusts the flag $\Psi^{t+1}=0$, locks the buyer clock, and switches to the seller's side.
    \item Clock Adjustment: The situations mentioned in the second step are situations where the buyer or seller satisfies the transaction. However, for the current auction clock, there may be no participants in the market who want to trade, so the auction clock needs to be adjusted. When $\Psi^t=0$ and no buyer can offer a bid, the auctioneer chooses a stepsize $\Theta^t$, which is a multiple of the minimum price interval, to adjust the buyer clock. Therefore, the auctioneer reduces the buyer clock as
	\begin{equation}
		C_B^{t+1} = C_B^t - \Theta^t.
	\end{equation}
	If the current auction is on seller side, i.e., $\Psi^t=1$ and no seller can offer a sell-bid, the current seller clock price should be raised as
    \begin{equation}
		C_S^{t+1} = C_S^{t} + \Theta^t.
	\end{equation}
	The default step size in the Vanilla DDA is set to the minimum price interval. Note that the auctioneer can also adjust this step size dynamically, so it can improve the efficiency of the auction via proper step size.
	
	After the adjustment of the auction clock, the auctioneer can also check whether the two clocks are cross. The auction ends at time slot $T\leftarrow t$ when the two clocks cross, i.e., $C_B^t < C_S^t$. At this time, the clear market price is set at the common crossing price $p^* = \kappa C_B^T + (1-\kappa) C_S^T$, where $\kappa\in[0,1]$ is the pricing coefficient.
    \end{enumerate}
    
    The utility of buyer $m$ can be expressed as $u_m = b_m^t - p^*$. Moreover, the utility of seller $n$ can be expressed as $u_n = p^* - a_n^t$.
	Thus, the total social welfare among the buyers and sellers can be expressed as
	\begin{equation}
	    SW = \sum_{m\in\mathcal{M}_B^T} u_m + \sum_{n\in\mathcal{N}_s^T} u_n.
	\end{equation}

	The design of the double Dutch auction mechanism has several desirable properties:
	\begin{itemize}
		\item Individual Rationality: Each buyer or seller achieves non-negative utility in the DDA mechanism. 
		\item Truthfulness: The payment of buyers and the revenue of sellers is determined by the common crossing price. Therefore, buyers and sellers do not have any other incentive to submit their buy-bids or sell-bids except their true valuations to the VR services.
		\item Budget balance: In the DDA, the buyers and the sellers trade with the common crossing price $p^*$ of buyer clock and seller clock. Thus, the auctioneer gains a non-negative utility as the total payment received from all winning bidders is higher than or equal to the total revenue paid to all winning sellers.
	\end{itemize}

	\subsection{Markov Decision Process of the DDA}
	
	A Markov Decision Process is defined by the tuple $<\mathcal{S}, \mathcal{A}, \mathcal{P}, \mathcal{R}>$, where $\mathcal{S}$ is the state space of the DDA, $\mathcal{A}$ is the action space of the auctioneer in the DDA, $\mathcal{P}$ is the state transition probabilities, and $\mathcal{R}$ is the reward.
	\subsubsection{State Space}
We define the state space at the current auction slot $t(t=1,2,\ldots,T)$ as a union of the side flag of auction market $\Psi^t$, the buyer clock $C_B^t$, the seller clock $C_S^t$, the size of winning buyer set $|\mathcal{M}_B^t|$, and the size of winning seller set $|\mathcal{N}_S^t|$, which is denoted as
	\begin{equation}
		S^t \triangleq \{\Psi^t,C_B^t,C_S^t, |\mathcal{M}_B^t|,|\mathcal{N}_S^t|\}.
	\end{equation}
	\subsubsection{Action Space}
	Action space of auctioneer includes the available step size $\Theta^t$ of the buyer clock and the seller clock at decision slot $t$, i.e., $a^t\triangleq \{ \Theta^t\}$.
	
	\subsubsection{Reward} 
    The reward of auctioneer in the DDA consists of two parts, the trade regret of buyers and sellers and the auction information exchange cost to broadcast the new auction clock to buyers or sellers, which can be expressed as
    \begin{align}
\begin{split}
r(S^t,a^t,S^{t+1})= \left \{
\begin{array}{ll}
    -e_m^t,                    & \Psi^t =0, b_m^t\geq C_B^t,\\
    -e_n^t,                    & \Psi^t =1, b_n^t\leq C_S^t,\\
    -p_a(N),   & \Psi^t =0, b_m^t< C_B^t,\\
    -p_a(M),   & \Psi^t =1, b_n^t> C_S^t,
\end{array}
\right.
\end{split}
\end{align}
where $p_a(X)$ is the auction information exchange cost function for $X$ participants in this market size with augmented Lagrangian methods, which can be expressed as follows:
\begin{equation}
    p_a(X) = cX, 
\end{equation}
and $c$ is the penalty of communication in auction.

	\subsubsection{Value Function}
	Given policy $\pi: \pi(a^t|S^t)$, the value function $V_\pi(S)$ of the state $S$ of the auctioneer, the expected discounting reward when starting in $S$ and following $\pi$ thereafter, can be expressed as
	\begin{equation}
		V_\pi(S) := \mathbb{E}_\pi\left[\sum_{t=0}^{T-1}\gamma^t r(S^t,a^t,S^{t+1})|S^0=S\right],
	\end{equation}
    where the expected value of a random variable given that the agent follows policy $\pi$ is denoted by $\mathbb{E}_\pi(\cdot)$ and the reward discount factor is denoted by $\gamma\in[0,1]$ used to reduce the weights as the time step increases.
	
\subsection{Evaluation and Improvement of Auctioneer Policy}
	
	The value function can be computed to help each agent formulate a better policy. For any arbitrary policy $\pi$ with a determined value function $V_\pi$, we can improve it through policy iteration. Using policy gradients, we aim to increase the likelihood of actions that will lead to higher expected rewards and reduce the likelihood of actions that will lead to lower expected rewards, until we obtain the optimal policy.
	To parameterize the policy and value function of the auctioneer, we use the neural network parameter $\vartheta$.
	We update the network parameters every time we train by randomly selecting samples from the replay buffer. We compute variance-reduced advantage-function estimators $A(S,a)$ employing a learning state-value function $V(S)$. Finally, we update the policy and value function of the auctioneer by maximizing the objective via stochastic gradient ascent following the proximal policy optimization update rules~\cite{schulman2017proximal}.
	\section{Simulation Results}
	\begin{figure}[t]
		\centering
		\includegraphics[width=0.8\linewidth]{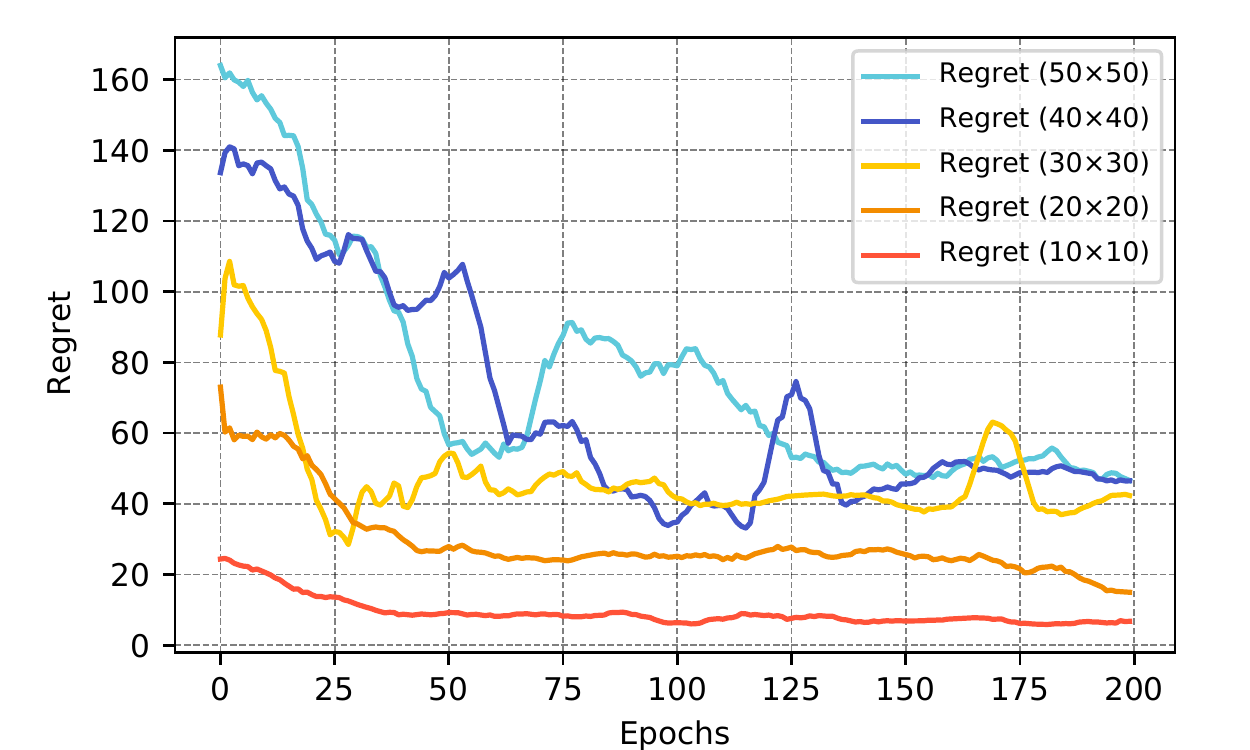}
		\caption{Convergence of the DRL-based DDA under different sizes of market.}
		\label{fig_regret}
	\end{figure}
	We consider a wireless edge-empowered Metaverse that consists of 10, 20, 30, 40, 50 VR users and 10, 20, 30, 40, 50 SPs. The average rotation speed of VR users ranges from 360$^{\circ}$ to 1080$^{\circ}$~\cite{Kelkkanen2020BitrateRO}. The required bitrate $R^{req}$ to support VR applications is set to 20 Mbps. The maximum transmits the power of SP to provide one VR service is set to 20 dBm. Thus the spectrum efficiency ranges from 1 to 3. The channel bandwidth of each resource block ranges from 0.1 to 0.2 MHz~\cite{chen2018virtual}. The resolution for HMD devices of VR users is 2160$\times$1200@90 fps, i.e., $D=2160\times 1200$ and $f=90$. The computation resource owned by SP is sampled from $f_n\in[2,4]$ GHz, and $\alpha$ is set to 0.01. The highest price $p^{max}$ is set to 100 and the lowest price $p^{min}$ is set to 1. The weight is set to $w_1 = w_2 = 0.5$. The minimum price interval is set to 1. The auction information exchange cost $c$ is set to 0.1.
\subsubsection{Baseline Methods}
    \begin{figure*}[t]
		\vspace{-0.38cm}
		\subfigure[Social welfare and Auction information exchange cost vs. Bitrates.]{
			\begin{minipage}[t]{0.325\linewidth}
				\includegraphics[width=1\linewidth]{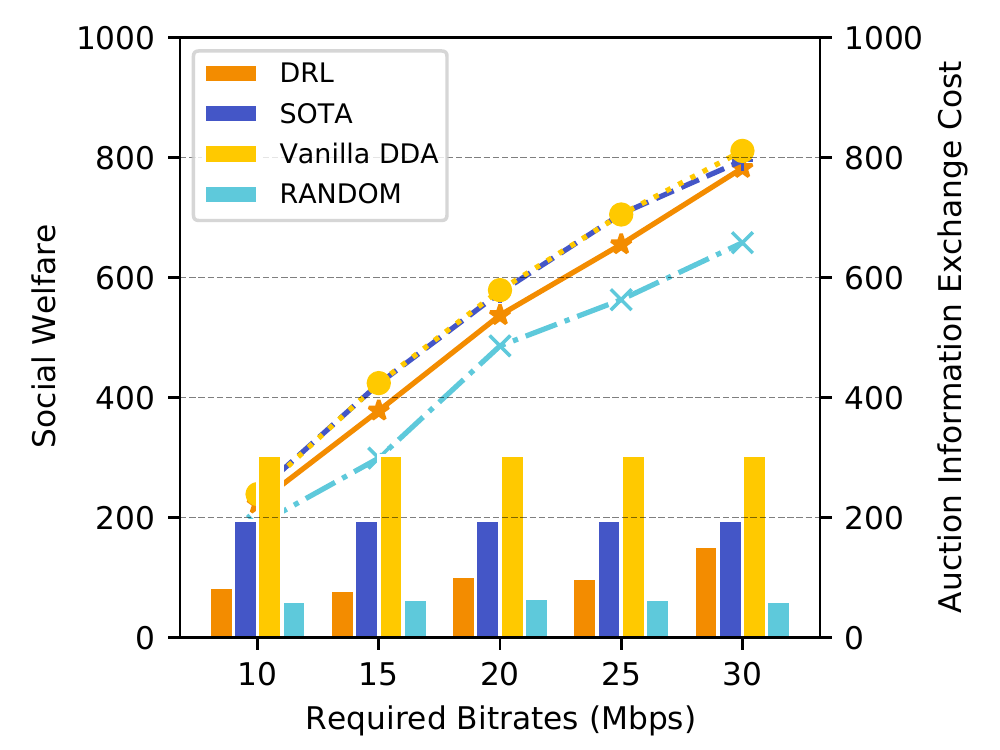}
			\end{minipage}%
		}%
		\subfigure[Social welfare vs. Sizes of market.]{
			\begin{minipage}[t]{0.325\linewidth}
				
				\includegraphics[width=1\linewidth]{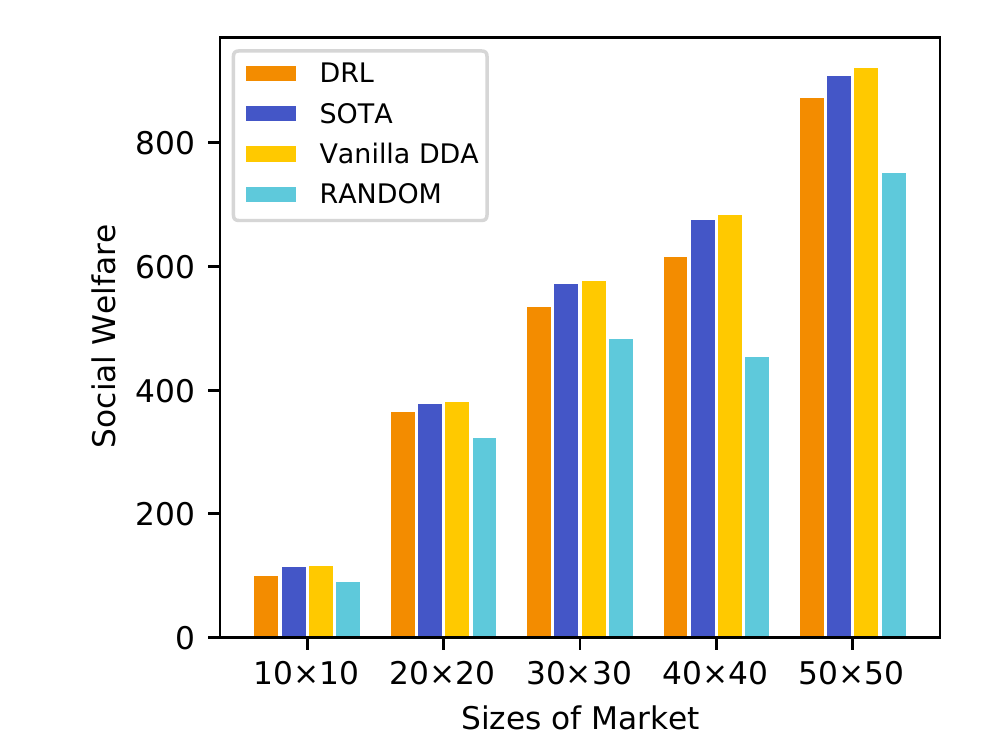}
			\end{minipage}%
		}%
		\subfigure[Auction information exchange cost vs. Sizes of market.]{
	\begin{minipage}[t]{0.325\linewidth}
				
				\includegraphics[width=1\linewidth]{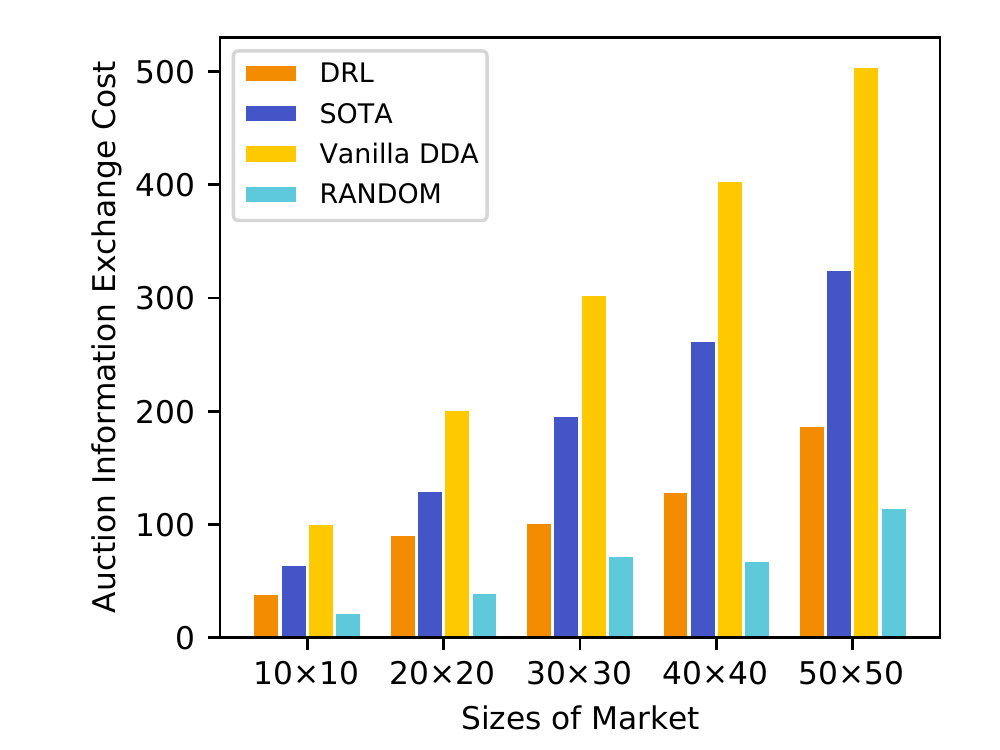}
			\end{minipage}%
		}%
		\centering
		\caption{Performance evaluation under different system parameters}
		\vspace{-0.38cm}
		\label{simulation}
	\end{figure*}
\begin{itemize}
\item Vanilla DDA: The auctioneer updates the auction clocks with the minimum price interval.
\item  State-of-the-art (SOTA) method~\cite{friedman2018double}: The auctioneer updates the auctions clocks with stepsize sampled from the Ornstein-Uhlenbeck process, which is large at the beginning and then converges to minimum price interval.
\item Random method: The auctioneer chooses the clock update stepsize from a uniform distribution.
\end{itemize}
	
	\subsection{Convergence Analysis}
	Firstly, we analyze the convergence performance of the proposed DRL-based DDA algorithm under different sizes of the VR service market. The sizes of VR service markets range from 10 buyers and 10 sellers (10$\times$10) to 50 buyers and 50 sellers ($50\times$50). As shown in Fig. 2, the DRL-based DDA can converge under various scales of the market. However, the speed to reach the convergence is different. For example, the DRL-based DDA converges faster in small-scale markets, which costs only about 30 epochs to achieve. However, in large-scale markets, such as 50$\times$50, it costs about 160 epochs to converge to the solution. Clearly, the DRL-based auctioneer needs to make more interactions with the market to learn participants' valuation for their VR services. Moreover, in larger markets, the DRL-based auctioneer's performance is less stable in its learning process, as shown by the yellow line representing 30$\times$30. The reason is that fitting the valuation of a large number of participants is not a trivial task.
	
    \subsection{Performance Evaluation under Different system parameters}
    
    Then, we compare the performance of the proposed DRL-based DDA under different system parameters, such as the required bitrate of VR services and sizes of the VR service market, with other baseline methods. Under the size of the market with 30$\times$30, Fig. 3(a) shows the trend of social welfare and auction information exchange cost, as the required bitrate of VR services increases. From Fig. 3(a), the social welfare in the market increases as the required bitrate increases. The reason is that as the required bitrates increase, the quality of perception experience of VR users increases, so their utility increases. Moreover, the auction information exchange cost of the DRL-based DDA is dynamic during the increasing of required bitrate, while others are static.
    
    In Fig. 3(b), we show the comparison with other baseline methods on achieved social welfare and auction information exchange cost under different sizes of markets. It can be observed that the SOTA and the Vanilla DDA methods achieve similar social welfare. This social welfare is also the maximum social welfare in the market as the Vanilla DDA is a regret-zero method. The social welfare achieved by the DRL-based DDA is about 5\% less than the maximum social welfare. However, the performance of the DRL-based DDA in social welfare is much better than the Random method, whose achieved social welfare is about 20\% less than that of the maximum social welfare. Moreover, in Fig. 3(c), we can observe that the auction information exchange cost of SOTA and Vanilla DDA is very large. Although the DRL-based DDA faces a slight loss in social welfare, its auction efficiency is much better than other baseline methods. The DRL-based DDA only incurs one-third of the auction information exchange cost of the Vanilla DDA and half of the SOTA methods.
    
    \section{Conclusion}
	In this paper, we have studied the economic system in the wireless edge-empowered Metaverse for VR services. We have proposed a learning-based incentive mechanism framework to evaluate and enhance the VR experience in the Metaverse efficiently. By using non-panoramic VR to reduce data traffic and computation consumption, we have applied SSIM and VMAF to formulate the experience of perception of VR users jointly. Then, to facilitate the trading of VR services, we have proposed the double Dutch auction to facilitate the matching and pricing processes between VR SPs and users asynchronously. To further enhance the efficiency of the DDA, we have trained the auctioneer in the DDA as the learning agent via DRL without prior auction knowledge. Finally, the experimental results demonstrate that our proposed economic framework for the Metaverse can achieve effective and efficient performance in terms of social welfare and auction information exchange cost in various system settings.

	\bibliographystyle{IEEEtran}
	\bibliography{conference_101719}
	
	%
	
	

	
	

\end{document}